\documentclass[11pt]{article}
\usepackage[a4paper,margin=0.8in]{geometry}
\usepackage{authblk}

\usepackage{amssymb}
\usepackage{graphicx}% Include figure files
\usepackage{dcolumn}% Align table columns on decimal point
\usepackage{bm}% bold math
\usepackage{amsmath}
\usepackage{txfonts}
\usepackage{amsfonts}
\usepackage[mathscr]{euscript}

\usepackage{booktabs}
\usepackage{tabularx}
\usepackage{tabulary}

\usepackage{multicol}
\usepackage{multirow}

\usepackage{array}

\usepackage{subfigure}

\usepackage{pdflscape}
\everymath{\displaystyle}
\DeclareMathOperator*{\argmax}{\arg\!\max}
\DeclareMathOperator*{\argmin}{\arg\!\min}

\title{A Bayesian fusion model for space-time reconstruction of finely resolved velocities in turbulent flows from low resolution measurements}

\author[1,2]{Linh Van Nguyen}
\author[1]{Jean-Philippe Laval}
\author[2]{Pierre Chainais}
\affil[1]{USTL, LML, CNRS UMR 8107, F-59650 Villeneuve d'Ascq, France}
\affil[2]{Ecole Centrale Lille, CRIStAL-CNRS UMR 9189, F-59650 Villeneuve d'Ascq, France}

\begin{document}
\maketitle
\begin{abstract}
The study of turbulent flows calls for measurements with high resolution both in space and in time. We propose a new approach to reconstruct High-Temporal-High-Spatial resolution velocity fields by combining two sources of information that are well-resolved either in space or in time, the Low-Temporal-High-Spatial (LTHS) and the High-Temporal-Low-Spatial (HTLS) resolution measurements. In the framework of co-conception between sensing and data post-processing, this work extensively investigates a Bayesian reconstruction approach using a simulated database. A Bayesian fusion model is developed to solve the inverse problem of data reconstruction. The model uses a Maximum A Posteriori estimate, which yields the most probable field knowing the measurements. The DNS of a wall-bounded turbulent flow at moderate Reynolds number is used to validate and assess the performances of the present approach. Low resolution measurements are subsampled in time and space from the fully resolved data. Reconstructed velocities are compared to the reference DNS to estimate the reconstruction errors. The model is compared to other conventional methods such as Linear Stochastic Estimation and cubic spline interpolation. Results show the superior accuracy of the proposed method in all configurations. Further investigations of model performances on various range of scales demonstrate its robustness.  Numerical experiments also permit to estimate the expected maximum information level corresponding to limitations of experimental instruments.
\end{abstract}

\section{\label{sec:Introduction}Introduction}
Turbulence, though governed by Navier-Stokes equations, is extremely hard to predict due to its spatiotemporally intermittency as well as three-dimensional and irregular properties. It is also a multi-scale phenomenon where a very wide range of scales from the largest eddies to Kolmogorov micro-scales co-exist and interact. Since the ratio between the largest and the smallest scales increases with Reynolds number as $ Re^{3/4} $, flows with high Reynolds are the most challenging. Wall bounded flows are particularly difficult to model due to the overlap of several scaling regions as a function of distance to the wall. Coherent structures in such flows can extend up to several boundary layer thickness. The modeling of such structures and scales therefore requires extremely detailed flow information in both space and time.

Despite a constant progress, none of the experimental techniques, even in academic researches, is capable of providing spatiotemporally resolved information in sufficiently wide spatial domains and for diverse flow conditions. Particle Image Velocimetry (PIV), the most advanced turbulence measurement technique, cannot measure space-time resolved velocities. Stereoscopic PIV measures three-component velocities at high spatial resolution and large field-of-view, but limited to a low acquisition rate compared to the flow dynamics. High repetition tomographic PIV and Time-resolved PIV (TrPIV) are improving but still limited to small volumes and low speed flows. Other point-measurement techniques such as Hot Wire Anemometry (HWA) measure the full temporal dynamics. However, the combination of these devices to get a better spatial resolution is not straightforward and remains intrusive.

Direct Numerical Simulation (DNS) can provide reliable and fully resolved velocities of turbulent flows. It simulates the flows by directly solving Navier-Stokes equations. The computational cost of such a numerical approach is very high since the number of simulated grid points increases as $Re^{9/4}$. DNS therefore can simulate flows with low to moderate Reynolds and simple geometries only.

To have fully resolved velocities, one idea is to measure and combine two types of complementary measurements in space and time: the Low-Temporal-High-Spatial resolution (LTHS) and the High-Temporal-Low-Spatial resolution (HTLS) measurements. One particular example of such an idea is presented in Ref.~\cite{coudert2011double}. This joint experiment provides a database of high Reynolds boundary layer flows. The data are provided by a stereoscopic PIV synchronized with a rake of HWA probes. PIV is with a large field-of-view and at a high spatial resolution but at low acquisition frequency. HWA measurements are at an extremely high temporal resolution, but the spatial discretization of the rake of probes is very coarse compared to Kolmogorov scales.

Various methods have been proposed to combine such measured data of turbulent flows to recover the maximum information level. Linear Stochastic Estimation (LSE) is the most common one. Its introduction into turbulence community dates back to the works by Adrian \cite{adrian1977role,adrian1979conditional} and has been further investigated later \cite{guezennec1989stochastic,adrian1992stochastic,ewing1999examination}. These works use LSE as a tool to extract coherent structures from the measurements. Later works proposed various extensions such as multi-time, nonlinear or higher-order LSE \cite{mokhasi2009predictive,durgesh2010multi,nguyen2010proper,meyer2014provide}. In these works, unknown velocities are reconstructed from measurements of other quantities such as pressure or shear-stress. LSE can be also linked to Proper Orthogonal Decomposition (POD) to reduce the order of reconstruction problems \cite{bonnet1994stochastic}. 

The idea of combining sparse velocity measurements to obtain fully-resolved fields has not been addressed until recently \cite{melnick2012experimental,tu2013integration}. In Ref.~\cite{melnick2012experimental}, 3D smoke intensity and 2D PIV measurements are combined using a POD-LSE model to get fully resolved 3D velocities of a flow over a flat plate. POD-LSE estimation model has been developed further \cite{tu2013integration} with a reconstruction scheme based on a multi-time LSE. Either a Kalman filter or a Kalman smoother is used depending on the problem as real time estimation or data post processing. The model is tested using TrPIV measurements of a bluff-body wake at a low Reynolds number. Sparse velocity measurements are virtually extracted from the high resolution ones, while original data are used to estimate reconstruction errors.

LSE suffers from critical limitations though extensively used. First, as a conditional average, LSE estimates a set of coefficients that associate the so-called conditional eddies to one flow pattern \cite{adrian1979conditional}. Using these coefficients to reconstruct all velocity fields, LSE fails to capture coherent structures and misleads physical interpretations when particular patterns exist. Second, the reconstructed structures are independent of event magnitudes \cite{moser1990statistical}. Reconstructed flows are associated with weak fluctuations only. Last, LSE as a low pass filter reconstructs large scales only and lose flow details even at measured positions.

\begin{figure}
\centering
	\begin{minipage}[b]{0.475\columnwidth}
		\includegraphics[width=\columnwidth]{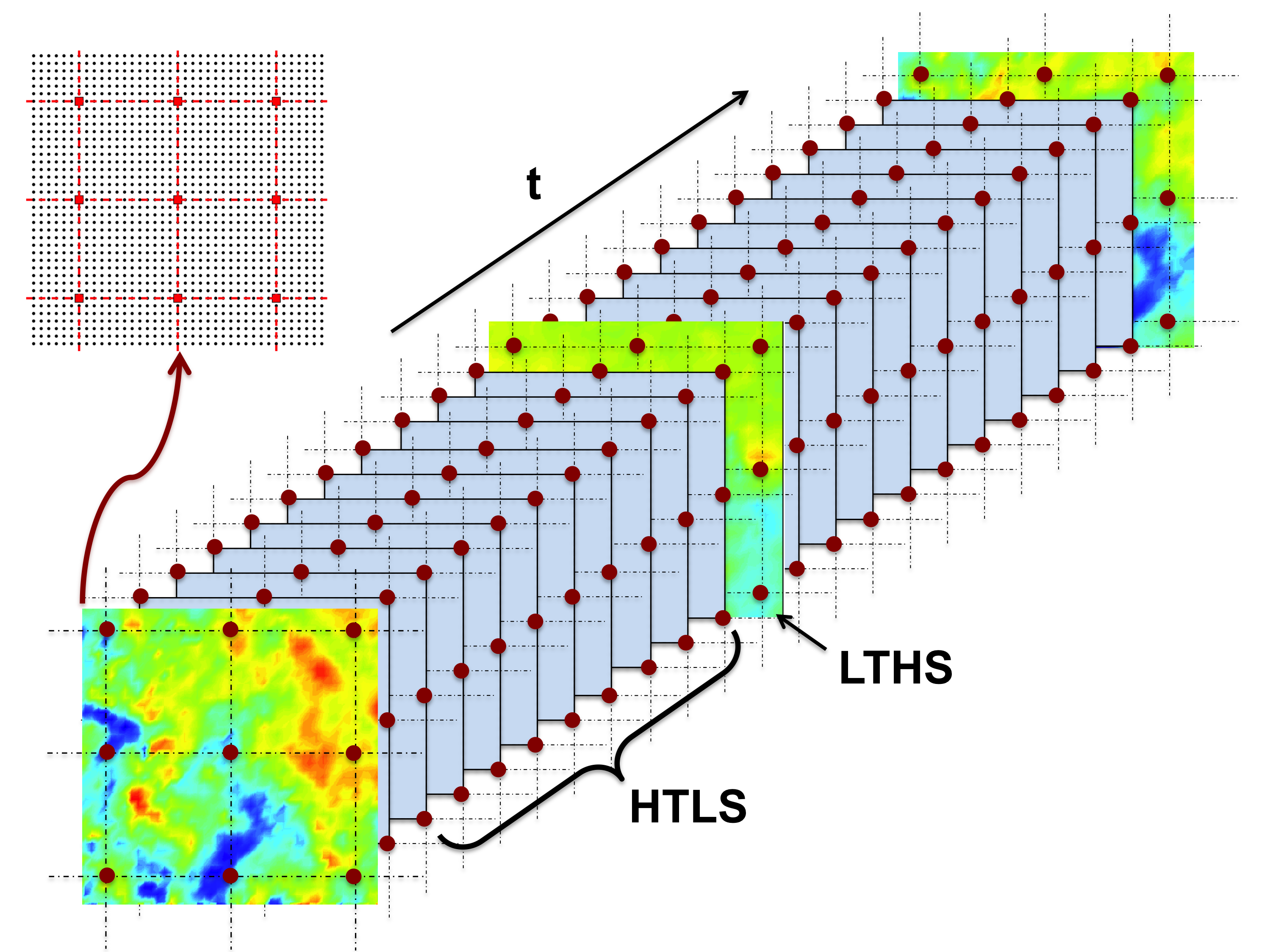}
		\caption{\label{fig:experiment} Sketch of the inverse problem, with the two sources of measurements: the LTHS (color images) and a coarse grid of HTLS (red dots among black ones of LTHS). The inverse problem of HTHS data reconstruction is to fill in the space-time data-cube.}
	\end{minipage}
	%\quad
	\hfill
	\begin{minipage}[b]{0.475\columnwidth}
		\includegraphics[width=\columnwidth]{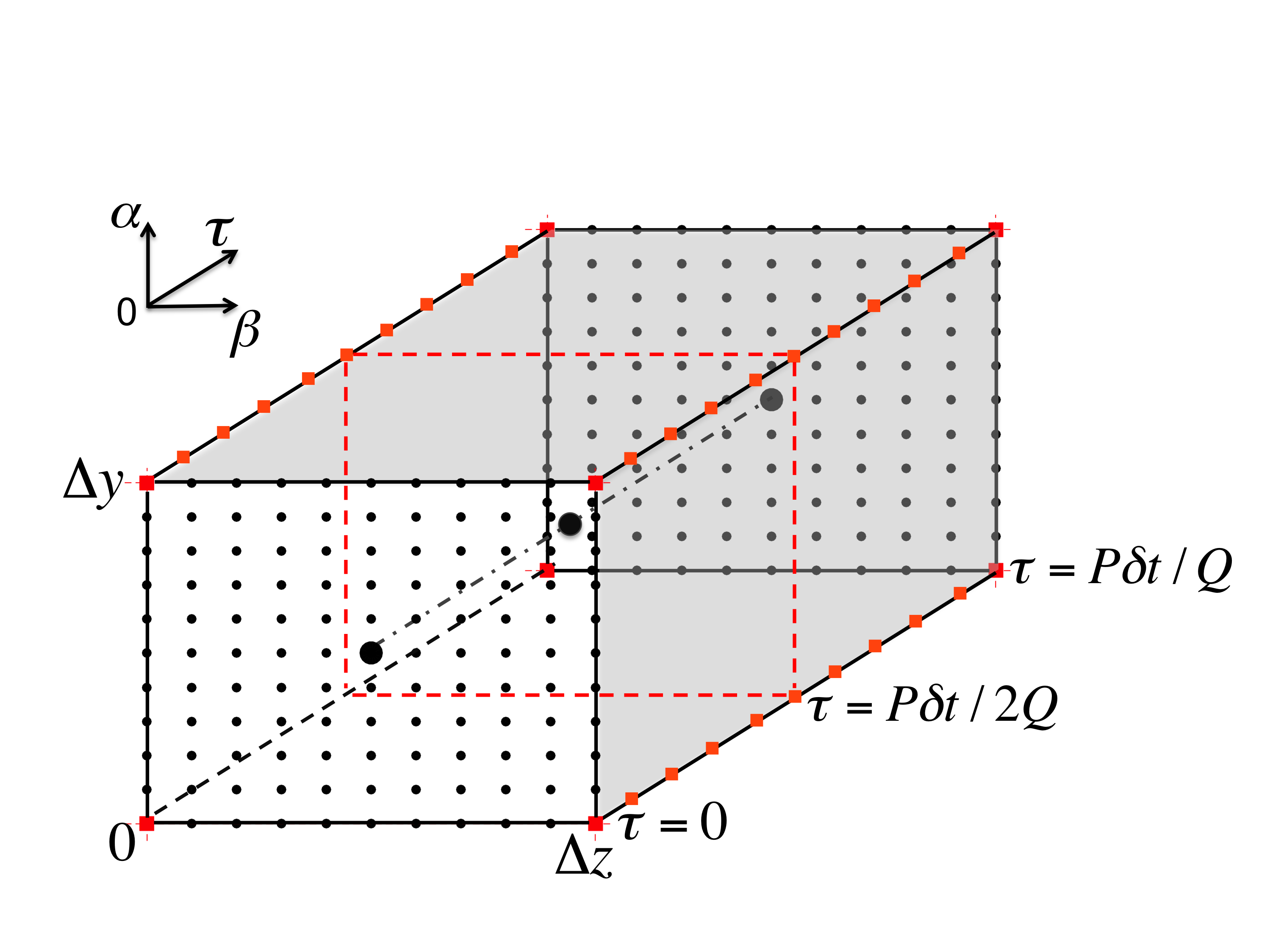}
		\caption{\label{fig:element_block} Sketch of an element block with local coordinates $ (\alpha,\beta,\tau) $. LTHS time steps are at $ \tau/\delta t=0 $ and $ \tau/\delta t=P/Q $. HTLS measurements are represented by red dots and LTHS measurements by black ones. }
	\end{minipage}
\end{figure}

The present work proposes a novel model to reconstruct the fully resolved HTHS velocities from HTLS and LTHS measurements. This model is based on a Bayesian inference framework using a Maximum A Posteriori (MAP) estimate \cite{kay1993fundamentals}. It is inspired by the multispectral image fusion problem with the limited resolution of image measurements in space-wavelength domains \cite{hardie2004map}. This framework has been discussed early in communication problems \cite{van1967detection,sage1971estimation} and is used more extensively in image processing, remote-sensing, and data fusion \cite{levitan1987maximum,ma2000simultaneous,challa2004bayesian,koks2003introduction,durrant2008multisensor,zhang2009noise,joshi2010map}.
The Bayesian fusion model takes benefit from both sources of information in space and time simultaneously by searching for the most probable flow for given measurements. Better performances are expected since space and time correlations are equally important. The model also recovers flow details inaccessible from single interpolations. By integrating directly the measurements, it proposes a compromise estimate such that detailed flow information close to the sensor positions are well preserved. This approach also overcomes the limitations of LSE, which acts as a low pass filter due to the mean square error minimization. 
To test the model, the DNS database of a turbulent wall-bounded flow is used. These space-time fully resolved data allow the  model optimization and validation. Sparse measurements of HTLS and LTHS are extracted from the full dataset, while the reference DNS data are used in the end to evaluate reconstruction errors. Performances are evaluated for various configurations with different subsampling ratios. 

The paper is organized as follows. Section \ref{sec:Bayesian_fusion_model} presents the Bayesian model using a MAP estimate. Model simplification and statistical parameters estimation are also discussed. Section \ref{sec:numerical_experiments} describes the DNS database used to test the model and also other reconstruction methods for comparison. Results for various configurations are presented. Conclusions and future works are in Section \ref{sec:conclusions}.

\section{\label{sec:Bayesian_fusion_model} Bayesian fusion model}
\begin{table}
	\centering
	\caption{\label{tab:summary_of_notations}Summary of notations.}
	\begin{tabulary}{\textwidth}{l p{11cm}} \toprule
	$ \boldsymbol{z} $  & $ NP- $dimensional vector of HTHS DNS data \\
	$ \boldsymbol{y} $  & $ MP- $dimensional vector of HTLS measurements \\
	$ \boldsymbol{x} $  & $ NQ- $dimensional vector of LTHS measurements \\
	$ N $  & number of spatial points in each HTHS/LTHS snapshot \\
	$ M $  & number of spatial points in each HTLS snapshot \\
	$ P $  & number of HTHS or HTLS snapshots \\
	$ Q $  & number of LTHS snapshots \\				
	$ \varmathbb{I}_t $ & 1D cubic spline interpolator in time\\
	$ \varmathbb{I}_s $ & 2D cubic spline interpolator in space\\
	$ \varmathbb{S}_t $ & subsampling in time from P to Q snapshots, $ \varmathbb{S}_t \boldsymbol{z} = \boldsymbol{x}$ \\
	$ \varmathbb{S}_s $ & subsampling in space from N to M points, $ \varmathbb{S}_s \boldsymbol{z} = \boldsymbol{y}$ \\
	$ \varmathbb{L}_s $ & 2D 5th-order least-square spline filter in space \\
	$ \varmathbb{L}_t $ & 1D 5th-order least-square spline filter in time \\
	$ \Delta\kappa $ & loss of kinetic energy computed using Eq.~(\ref{eq:RMS_losses}) \\	
	$ \epsilon $ & normalized Root Mean Square Error (NRMSE) defined in Eq.~(\ref{eq:NRMSE}) \\
	$ |.| $ & determinant of a matrix \\
	$\boldsymbol{u}^T$ & transpose of vector $ \boldsymbol{u} $ \\	
	$\argmax_{\boldsymbol{z}}(.)$ & argument of the maximum,  which is $ \boldsymbol{z} $ for which the function attains its maximum\\		
	$ \Arrowvert \boldsymbol{u} \Arrowvert^2_{\Sigma} $ & square of a Mahalanobis distance: $ \Arrowvert \boldsymbol{u} \Arrowvert^2_{\Sigma} = \boldsymbol{u}^T \Sigma^{-1} \boldsymbol{u}$\\
	$ \Arrowvert \boldsymbol{u} \Arrowvert^2_2 $ & square of a Euclidean distance: $ \Arrowvert \boldsymbol{u} \Arrowvert^2_2 = \boldsymbol{u}^T \boldsymbol{u}$\\	
	$\boldsymbol{n} \sim \mathscr{N}(0,\Sigma_{\boldsymbol{n}}) $ & Gaussian noise vector of zero-mean and covariance matrix $ \Sigma_{\boldsymbol{n}} $\\		
	$\mathscr{N}(\boldsymbol{u}|\boldsymbol{\mu}_{\boldsymbol{u}},\Sigma_{\boldsymbol{n}}) $ & Gaussian distribution of variable $ \boldsymbol{u} $ that takes the mean $\boldsymbol{\mu}_{\boldsymbol{u}} $ and fluctuates due to $ \boldsymbol{n} $ of covariance $ \Sigma_{\boldsymbol{n}} $\\
	$\mathscr{N}(\boldsymbol{u}|\boldsymbol{v})$ & distribution of $ \boldsymbol{u} $ knowing (or conditioning on) $ \boldsymbol{v} $\\	\bottomrule
	\end{tabulary}
\end{table}
\subsection{Bayesian model}
Let $ \boldsymbol{x}$ and $ \boldsymbol{y}$ denote LTHS and HTLS measurements, and $ \boldsymbol{z}$ denote HTHS data to reconstruct. $ \boldsymbol{z}$, $ \boldsymbol{x}$ and $ \boldsymbol{y}$ are random, zero-mean vectors of size $ NP \times 1$, $ NQ \times 1$ and $MP \times 1$ respectively.  $ N $ and $ M $ are numbers of spatial points in each snapshot, while $ P $ and $ Q $ are numbers of snapshots. The present work is a challenging inverse problem since we consider  $ M \ll N $ and $ Q \ll P $.
Let the subscript ``s'' denote operators performing in space, and ``t'' be those in time; $ \varmathbb{I} $ is an interpolator; $ \varmathbb{S} $ is for subsampling; $ \varmathbb{L} $ is a Low-Pass Filter (LPF). The cubic spline interpolation either 1D or 2D is used as $ \varmathbb{I} $ for its state-of-the-art interpolation results and finite support \cite{unser1999splines,thevenaz2000interpolation}. $ \varmathbb{L} $ is a 5th-order least-square spline filter \cite{reinsch1967smoothing,cook1981smoothing} for its sharp cutoff response to better separate large scales from small scales. Table~\ref{tab:summary_of_notations} lists all notations used in this paper.

Given sparse measurements of either $ \boldsymbol{y} $ in space or $ \boldsymbol{x} $ in time, the fully resolved vector $ \boldsymbol{z} $ can be reconstructed by single interpolations. The 1D time interpolation goes from $ NQ $ to $ NP $ dimensional space, i.e. $ \boldsymbol{x} \longmapsto \hat{\boldsymbol{z}}=\varmathbb{I}_t \boldsymbol{x} $, while the 2D space interpolation goes from  $ MP$ to $ NP $ dimensional space, i.e. $ \boldsymbol{y} \longmapsto \hat{\boldsymbol{z}}=\varmathbb{I}_s \boldsymbol{y}$. Let $ NP $ dimensional vectors $ \boldsymbol{h}_s $ and $ \boldsymbol{h}_t $ denote the information that cannot be recovered by simple interpolations; $\boldsymbol{z} $ can be modeled in two ways:
\begin{align}
	\boldsymbol{z} &= \varmathbb{I}_t\boldsymbol{x}+\boldsymbol{h}_t \label{eq:Bayes11}\\
	\boldsymbol{z} &= \varmathbb{I}_s\boldsymbol{y}+\boldsymbol{h}_s \label{eq:Bayes12}	
\end{align}
Missing information $ \boldsymbol{h}_t $ and $ \boldsymbol{h}_s $ essentially feature small scales. Using either $ \boldsymbol{x} $ or $ \boldsymbol{y} $, it is not possible to estimate $ \boldsymbol{h}_t $ and $ \boldsymbol{h}_s $. The idea of Bayesian fusion is to combine the two models by using $ \varmathbb{I}_t\boldsymbol{x} $ in (\ref{eq:Bayes11}) to estimate the unknown $ \boldsymbol{h}_s $ in ~(\ref{eq:Bayes12}) and vice-versa.

Let $\mathscr{N}(\boldsymbol{u}|\boldsymbol{\mu}_{\boldsymbol{u}},\Sigma_{\boldsymbol{u}}) $ denote the multivariate Gaussian distribution of a $ NP $ dimensional random vector $ \boldsymbol{u} $ with mean value $ \boldsymbol{\mu}_{\boldsymbol{u}} $ and covariance matrix $\Sigma_{\boldsymbol{u}}$.
%varying due to a random Gaussian noise $ \boldsymbol{n} $.
The $ NP \times NP $ matrix is the expectation of $ (\boldsymbol{u}-\boldsymbol{\mu}_{\boldsymbol{u}} )(\boldsymbol{u}-\boldsymbol{\mu}_{\boldsymbol{u}} )^T $. 
The probability density function (pdf) of $ \boldsymbol{u} $ with a multivariate Gaussian distribution $ \mathscr{N}(\boldsymbol{u}|\boldsymbol{\mu}_{\boldsymbol{u}},\Sigma_{\boldsymbol{u}}) $ is:
\begin{equation}
	p(\boldsymbol{u})=\frac{1}{(2\pi)^{NP/2}|\Sigma_{\boldsymbol{u}}|^{1/2}} e^{\displaystyle \Arrowvert \boldsymbol{u}-\boldsymbol{\mu}_{\boldsymbol{u}} \Arrowvert^2_{\Sigma_{\boldsymbol{u}}} }
	\label{eq:Bayes4}
\end{equation} 
where $ |.| $ denotes the matrix determinant, and $\Arrowvert \boldsymbol{u}-\boldsymbol{\mu}_{\boldsymbol{u}} \Arrowvert^2_{\Sigma_{\boldsymbol{u}}}$ is the Mahalanobis distance:
\begin{equation}
 \Arrowvert \boldsymbol{u}-\boldsymbol{\mu}_{\boldsymbol{u}} \Arrowvert^2_{\Sigma_{\boldsymbol{n}}}=\left(\boldsymbol{u}-\boldsymbol{\mu}_{\boldsymbol{u}}\right)^T \Sigma_{\boldsymbol{n}}^{-1}\left( \boldsymbol{u}-\boldsymbol{\mu}_{\boldsymbol{u}}\right)
\end{equation}
Let assume that $ \varmathbb{I}_t\boldsymbol{x} $ and $ \boldsymbol{h}_t $ are approximately independent; $ \varmathbb{I}_t\boldsymbol{x} $ captures temporal large scales of $ \boldsymbol{x} $. Similarly, $ \varmathbb{I}_s\boldsymbol{y} $ and $ \boldsymbol{h}_s $ are assumed to be approximately independent. Due to subsampling, aliasing terms are also present in each pairs of $ (\varmathbb{I}_t\boldsymbol{x}, \boldsymbol{h}_t) $ and $ (\varmathbb{I}_s\boldsymbol{y}, \boldsymbol{h}_s) $. Assume also that $ \boldsymbol{h}_t $ and $ \boldsymbol{h}_s $ are zero mean Gaussian noises, i.e. $ \boldsymbol{h}_t \sim \mathscr{N} (0,\Sigma_{\boldsymbol{h}_t}) $ and $ \boldsymbol{h}_s \sim \mathscr{N} (0,\Sigma_{\boldsymbol{h}_t}) $. Pdfs of these unknowns are modeled as:
\begin{equation}
	p(\boldsymbol{h}_t)=\frac{1}{(2\pi)^{NP/2}|\Sigma_{\boldsymbol{h}_t}|^{1/2}} e^{\displaystyle \Arrowvert \boldsymbol{h}_t \Arrowvert^2_{\Sigma_{\boldsymbol{h}_t}} }
	\label{eq:Bayes5}
\end{equation} 
and similarly for $ p(\boldsymbol{h}_s) $. Posterior distributions of $ \boldsymbol{z} $ knowing either $ \boldsymbol{x} $ or $ \boldsymbol{y} $ are then modeled as:
\begin{align}
	\mathscr{N}(\boldsymbol{z}|\boldsymbol{x}) &\sim \mathscr{N}(\boldsymbol{z}|\varmathbb{I}_t\boldsymbol{x},\Sigma_{\boldsymbol{h}_t}) \label{eq:Bayes21} \\
	\mathscr{N}(\boldsymbol{z}|\boldsymbol{y}) &\sim \mathscr{N}(\boldsymbol{z}|\varmathbb{I}_s\boldsymbol{y},\Sigma_{\boldsymbol{h}_s}) \label{eq:Bayes22}	
\end{align}
where $ \mathscr{N}(\boldsymbol{z}|\boldsymbol{x}) $, resp. $ \mathscr{N}(\boldsymbol{z}|\boldsymbol{x}) $, is the posterior distribution of $ \boldsymbol{z} $ knowing $ \boldsymbol{x} $, resp. knowing $ \boldsymbol{y} $.% and similarly is $ \mathscr{N}(\boldsymbol{z}|\boldsymbol{y}) $.

\subsection{\label{subsec:MAP_estimation} MAP estimation}
The present Bayesian model aims to build an estimate of $ \boldsymbol{z} $ given $ \boldsymbol{x} $ and $ \boldsymbol{y} $ using the probability models (\ref{eq:Bayes21}) and (\ref{eq:Bayes22}). The model uses a MAP estimate to search the most probable $ \hat{\boldsymbol{z}} $ given $ \boldsymbol{x}$ and $ \boldsymbol{y} $ such that $ \hat{\boldsymbol{z}} $ maximizes the posterior pdf $ p(\boldsymbol{z}|\boldsymbol{x},\boldsymbol{y}) $:
\begin{equation}
 \hat{\boldsymbol{z}}=\argmax_{\boldsymbol{z}} \:  p \left(\boldsymbol{z}|\boldsymbol{x},\boldsymbol{y}\right)
 \label{eq:MAP1}
\end{equation}
Using Bayesian rules \cite{kendall1987kendall}, one has:
\begin{equation}
p\left(\boldsymbol{z}|\boldsymbol{x},\boldsymbol{y}\right) \propto p\left(\boldsymbol{x},\boldsymbol{y}|\boldsymbol{z} \right)p(\boldsymbol{z})
 \label{eq:MAP2}
\end{equation}
Assuming that $ \boldsymbol{x} $ and $ \boldsymbol{y} $ are independent conditioned on $ \boldsymbol{z} $, Eq.~(\ref{eq:MAP2}) becomes:
\begin{equation}
p\left(\boldsymbol{z}|\boldsymbol{x},\boldsymbol{y}\right) \propto p(\boldsymbol{x}|\boldsymbol{z})p(\boldsymbol{y}|\boldsymbol{z})p(\boldsymbol{z})
\label{eq:MAP3}
\end{equation}
In Eq.~(\ref{eq:MAP3}), the likelihood functions $ p(\boldsymbol{x}|\boldsymbol{z}) $, $ p(\boldsymbol{y}|\boldsymbol{z}) $ and the prior pdf $ p(\boldsymbol{z}) $ appear, while only the posterior probabilities $ p(\boldsymbol{z}|\boldsymbol{x}) $ and $ p(\boldsymbol{z}|\boldsymbol{y}) $ are available in ~(\ref{eq:Bayes21}) and ~(\ref{eq:Bayes22}). 

To complete the model, the likelihood functions can be expressed in term of the posterior pdfs and prior of $ \boldsymbol{z} $ using Bayesian rules. Section 2.3.3 in Ref.~\cite{bishop2006pattern} introduces an alternative way to estimate these functions from posterior pdfs using a linear Gaussian model. Various tests of different Gaussian priors $ p(\boldsymbol{z}) $ lead to the use of a noninformative prior. This prior, referred also as vague or flat prior, assumes that all the values of $ \boldsymbol{z} $ are equally likely \cite{gelman2014bayesian}. The estimation of $ \hat{\boldsymbol{z}} $ is now solely based on the measurements and not influenced by external information. The prior distribution therefore has no influence on the posterior pdfs.

With the assumption of a noninformative prior, $ p(\boldsymbol{z}) $ is constant. Using Bayes rules, the relation between the likelihood function and the posterior pdf is:
\begin{equation}
p\left(\boldsymbol{z}|\boldsymbol{x}\right) \propto p\left(\boldsymbol{x}|\boldsymbol{z}\right)p(\boldsymbol{z})
\end{equation}
Since $ p(\boldsymbol{z}) $ is replaced by a constant, one gets $p\left(\boldsymbol{x}|\boldsymbol{z}\right) \propto p\left(\boldsymbol{z}|\boldsymbol{x}\right)$. Similarly, $p\left(\boldsymbol{y}|\boldsymbol{z}\right) \propto p\left(\boldsymbol{z}|\boldsymbol{y}\right)$. Eq~(\ref{eq:MAP3}) becomes:
\begin{equation}
p\left(\boldsymbol{z}|\boldsymbol{x},\boldsymbol{y}\right) \propto p(\boldsymbol{z}|\boldsymbol{x})p(\boldsymbol{z}|\boldsymbol{y})
\label{eq:MAP4}
\end{equation}
The MAP estimation is: 
\begin{equation}
	\hat{\boldsymbol{z}}= \argmax_{\boldsymbol{z}} \: p(\boldsymbol{z}|\boldsymbol{x})p(\boldsymbol{z}|\boldsymbol{y})\
\label{eq:MAP_5} 
\end{equation}
Logarithms of $ p\left(\boldsymbol{z}|\boldsymbol{x}\right) $ and $ p\left(\boldsymbol{z}|\boldsymbol{y}\right) $ are:
\begin{align}
	-\ln p(\boldsymbol{z}|\boldsymbol{x}) &= \frac{1}{2}\Arrowvert \boldsymbol{z}-\varmathbb{I}_t\boldsymbol{x} \Arrowvert^2_{\Sigma_{\boldsymbol{h}_t}} +C_1 \label{eq:MAP61}\\
	-\ln p(\boldsymbol{z}|\boldsymbol{y}) &= \frac{1}{2} \Arrowvert \boldsymbol{z}-\varmathbb{I}_s\boldsymbol{y} \Arrowvert^2_{\Sigma_{\boldsymbol{h}_s}} +C_2 \label{eq:MAP62}
\end{align}
where $ C_1 $ and $ C_2 $ are independent of $ \boldsymbol{x} $, $ \boldsymbol{y} $ and $ \boldsymbol{z} $. Solving (\ref{eq:MAP_5}) is equivalent to minimize the cost function:
\begin{equation}
	C(\boldsymbol{z})=\frac{1}{2} \Arrowvert \boldsymbol{z}-\varmathbb{I}_t\boldsymbol{x} \Arrowvert^2_{\Sigma_{\boldsymbol{h}_t}} + \frac{1}{2}\Arrowvert \boldsymbol{z}-\varmathbb{I}_s\boldsymbol{y} \Arrowvert^2_{\Sigma_{\boldsymbol{h}_s}}
\label{eq:MAP7}
\end{equation}
Computing the gradient of $ C(\boldsymbol{z}) $  and setting to zero:
\begin{equation}
	\frac{\partial C(\boldsymbol{z})}{\partial \boldsymbol{z}}= \Sigma^{-1}_{\boldsymbol{h}_s} \left( \boldsymbol{z}-\varmathbb{I}_s\boldsymbol{y} \right) + \Sigma^{-1}_{\boldsymbol{h}_t} \left( \boldsymbol{z}-\varmathbb{I}_t\boldsymbol{x} \right) = 0
\label{eq:MAP8}	
\end{equation}
the solution to the optimization problem (\ref{eq:MAP1}) is:
\begin{equation}
	\hat{\boldsymbol{z}} = \left( \Sigma^{-1}_{\boldsymbol{h}_t} + \Sigma^{-1}_{\boldsymbol{h}_s} \right)^{-1} \left(\Sigma^{-1}_{\boldsymbol{h}_s}\varmathbb{I}_s\boldsymbol{y}+\Sigma^{-1}_{\boldsymbol{h}_t}\varmathbb{I}_t\boldsymbol{x}\right)
\label{eq:MAP9}
\end{equation}
Applying the matrix inversion lemma \cite{kay1993fundamentals}: 
\begin{equation}
	(A+BD^{-1}C)^{-1}=A^{-1}-A^{-1}B(D+CA^{-1}B)^{-1}CA^{-1}
\label{eq:inversion_lemma}	
\end{equation}
Eq.~(\ref{eq:MAP9}) can be rewritten as:
\begin{equation}
	\hat{\boldsymbol{z}} = \left( \Sigma_{\boldsymbol{h}_t} + \Sigma_{\boldsymbol{h}_s} \right)^{-1} \left(\Sigma_{\boldsymbol{h}_t}\varmathbb{I}_s\boldsymbol{y}+\Sigma_{\boldsymbol{h}_s}\varmathbb{I}_t\boldsymbol{x}\right)
\label{eq:MAP10}
\end{equation}
Eq.~(\ref{eq:MAP10}) is the final full form of the proposed Bayesian fusion model using a MAP estimate and assuming a noninformative prior of $ \boldsymbol{z} $. Variance matrices $ \Sigma_{\boldsymbol{h}_t} $ and $ \Sigma_{\boldsymbol{h}_s} $ are parameters to be estimated. 

\subsection{\label{subsec:simplified_fusion_model} Model simplification}
Though providing the full theoretical estimate of $ \hat{\boldsymbol{z}} $, Eq.~(\ref{eq:MAP10}) is impractical to use as is for several reasons. The full covariance matrices $ \Sigma_{\boldsymbol{h}_t} $ and $ \Sigma_{\boldsymbol{h}_s} $, representing all the sources of correlations in space and time, cannot be estimated from only the measurements $ \boldsymbol{x} $ and $ \boldsymbol{y}$. This is because the unknown $ \boldsymbol{h}_t $ and $ \boldsymbol{h}_s $ are only accessible at the measured positions in space and time. Also, the covariance matrices of size $ NP \times NP $ are very large, making them very difficult to accurately estimate and to inverse. Additional assumptions on the shape of $ \Sigma_{\boldsymbol{h}_t} $ and $ \Sigma_{\boldsymbol{h}_s} $ are necessary. 

A common and simple approach is to assume diagonal covariance matrices. This implies the independence between all elements of $ \boldsymbol{h}_t $ and $ \boldsymbol{h}_s $. The simplified version of Eq.~(\ref{eq:MAP10}) becomes a point-wise formula:
\begin{equation}	
	\hat{\boldsymbol{z}}(i)=\frac{\boldsymbol{\sigma}^2_{\boldsymbol{h}_s}(i)}{\boldsymbol{\sigma}^2_{\boldsymbol{h}_s}(i)+\boldsymbol{\sigma}^2_{\boldsymbol{h}_t}(i)}\varmathbb{I}_t\boldsymbol{x}(i)+\frac{\boldsymbol{\sigma}^2_{\boldsymbol{h}_t}(i)}{\boldsymbol{\sigma}^2_{\boldsymbol{h}_s}(i)+\boldsymbol{\sigma}^2_{\boldsymbol{h}_t}(i)}\varmathbb{I}_s\boldsymbol{y}(i)
\label{eq:MAP_simplified}	
\end{equation}
where  $ i(t,s) $ is the index of each point in time $ (t) $ and space $ (s) $. The variances $ \boldsymbol{\sigma}^2_{\boldsymbol{h}_s} $ and $ \boldsymbol{\sigma}^2_{\boldsymbol{h}_t} $ are functions of each position in space and time. Their estimation is detailed in next section. Tehn Eq.~(\ref{eq:MAP_simplified}) will be used to reconstruct HTHS data. As a weighted average, it proposes a compromise estimate from the measurements. With a symmetrical form in space and time, the model uses information from both measurements to correct large scales reconstruction and recover certain information at smaller scales. 

\subsection{\label{subsec:statistical_parameters_estimation} Statistical parameters estimation}
Let $Z$, $ X $, $ Y $,$ H_t $, $ H_s $, $ \Gamma_{\boldsymbol{h}_t} $ and $ \Gamma_{\boldsymbol{h}_s} $ be the (time,space) matrix forms of $ \boldsymbol{z} $, $ \boldsymbol{x} $, $ \boldsymbol{y} $, $ \boldsymbol{h}_t $, $ \boldsymbol{h}_s $, $\boldsymbol{\sigma}^2_{\boldsymbol{h}_t} $ and $\boldsymbol{\sigma}^2_{\boldsymbol{h}_s} $ respectively.  $Z$, $ H_t $, $ H_s $, $ \Gamma_{\boldsymbol{h}_t} $ and $ \Gamma_{\boldsymbol{h}_s} $ are of size $ P\times N $, while $ X $ and $ Y $ are of size $ Q\times N $ and $ P\times M $. $ \Gamma_{\boldsymbol{h}_t} $ and $ \Gamma_{\boldsymbol{h}_s} $ are matrices of empirical variances, which are functions of time and space $ (t,s) $. 

Variance matrices are estimated from $ H_t $ and $ H_s $, which are available at the measurements positions only. We use:
\begin{align}
	\varmathbb{S}_tH_s &= X- \varmathbb{I}_s\varmathbb{S}_sX\\
	\varmathbb{S}_sH_t &= Y- \varmathbb{I}_t\varmathbb{S}_tY
\end{align}
where $ \varmathbb{S}_t $ subsamples in time from $ P $ to $ Q $ time steps, and $ \varmathbb{S}_s $ subsamples in space from $ N $ to $ M $ points. These Q instants and M positions are the same as for LTHS and HTLS measurements. Since the flow is approximately stationary and spatial interpolation is independent of time, $ \Gamma_{\boldsymbol{h}_s} (t,s) $ becomes $ \Gamma_{\boldsymbol{h}_s} (s) $, a function of spatial locations only. These variances are estimated by averaging over all time steps:
\begin{equation}
	\Gamma_{\boldsymbol{h}_s} (s) = \frac{1}{Q}\sum_{t=1}^{Q} \left( X(t,s)- \varmathbb{I}_s\varmathbb{S}_sX(t,s) \right)^2 
	\label{eq:variance_estimation_1}
\end{equation}
Variance in $ \Gamma_{\boldsymbol{h}_t}$ is a function of distances $ \tau $ to the previous LTHS time step only, where $ \tau/\delta t=0,1,2,...,P/Q $, and $ \delta t $ is the time lag between two consecutive HTHS time steps. $ \Gamma_{\boldsymbol{h}_t} $ becomes a function of space and $ \tau $ only, i.e. $ \Gamma_{\boldsymbol{h}_t} (\tau,s) $. It is estimated by averaging over Q blocks (of $ P/Q $ snapshots) bounded by two consecutive LTHS instants: 
\begin{equation}
	\Gamma_{\boldsymbol{h}_t} (\tau,s) = \frac{1}{Q}\sum_{t_s} \left( Y(t_s,s)- \varmathbb{I}_t\varmathbb{S}_tY(t_s,s) \right)^2
	\label{eq:variance_estimation_2}
\end{equation}
where $ t_s/\delta t=\tau/\delta t,\tau/\delta t+P/Q,\tau/\delta t+2P/Q,...,\tau/\delta t+(Q-1)P/Q $. Since the flow is approximately homogeneous in spanwise direction, $ \Gamma_{\boldsymbol{h}_s} (s) $ and $ \Gamma_{\boldsymbol{h}_t} (\tau,s) $ are also averaged over all blocks defined by the four neighboring HTLS measurements, see Fig.~\ref{fig:element_block}. The variances are then functions of only vertical positions and relative distances to the four closest HTLS sensors. These estimated variances are rearranged into a vector form $\boldsymbol{\sigma}^2_{\boldsymbol{h}_t}(i) $ and $\boldsymbol{\sigma}^2_{\boldsymbol{h}_s}(i) $ to complete the fusion model using the simplified formula in Eq.~(\ref{eq:MAP_simplified}).

% % % % % % % % % % % % % % % % % % % % % % % % % % % % % % % % % % % % % % % %
% % % % % % % % % % % % % % % % % % RESULTS % %% % % % % % % % % % % % % % % % % % % % %% % % % % % % % % % % % % % % % % % % %% % % % % % % % % % % % % % % % % % % % % % % % % % % % % % % % % % % % % % % % % % % % % % % % % % % % % % %
\begin{landscape}
\begin{table}
\caption{\label{tab:results}
NRMSEs of all scales reconstruction errors for various cases. The subsampling ratios of HTLS measurements are $ \sqrt{N/M} $ and equal in both spatial directions. The ratios of LTHS measurements in time are $ P/Q $. The equivalent spacing in spanwise direction is normalized by half channel height as $ \Delta z/H $ and the spacing in time is $\Delta t$. The normalized energy losses in space $\Delta\kappa_s$ and in time $\Delta\kappa_t$ are defined in Eq.~(\ref{eq:RMS_losses}). $\overline{\epsilon}$ and $\epsilon_{max}$ are the mean and max NRMSE defined in Eq.~(\ref{eq:NRMSE}). $\overline{\epsilon}$ is averaged over all space-time positions in the outer region $ y/H \in [0.25,1.75] $, while $\epsilon_{max}$ is computed for one of the most difficult position in space and time (the most remote from all nearby measurements). The smallest errors in each cases are boldfaced.}
\vspace{.5cm}
\centering
\begin{tabulary}{0.8\textwidth}{|c|cc|cc|cc|cccc|cccc|}
\hline \multicolumn{1}{|c|}{}&\multicolumn{2}{c|}{Subsampling ratios}&\multicolumn{2}{c|}{Spacings}&\multicolumn{2}{c|}{Energy loss}&\multicolumn{4}{c|}{$\overline{\epsilon}$}&\multicolumn{4}{c|}{$\epsilon_{max}$}\\ \hline
\rule{0pt}{2.5ex} {Case} & $\sqrt{N/M}$ & $P/Q$ & {$\Delta z/H$} & {$\Delta t \: (s)$} &$\Delta\kappa_s(\%)$ &$\Delta\kappa_t (\%)$& {$\varmathbb{I}_s\boldsymbol{y}$} & {$\varmathbb{I}_t\boldsymbol{x}$} & {LSE} & {Fusion} & {$\varmathbb{I}_s\boldsymbol{y}$} & {$\varmathbb{I}_t\boldsymbol{x}$} & {LSE} & {Fusion}\\
\hline
1  & 05  & 10 & 0.05  & 0.25 & 0.29 & 4.70  & 0.14 & 0.32 & 0.25 &  {\textbf{0.12}} & {\textbf{0.16}} & 0.56 & 0.33  &  {\textbf{0.16}} \\
2  & 05  & 20 & 0.05  & 0.50 & 0.29 & 13.12 & 0.14 & 0.54  & 0.38 &  {\textbf{0.13}} & {\textbf{0.16}} & 0.89 & 0.49  &  {\textbf{0.16}}\\
3  & 10  & 04 & 0.11  & 0.10 & 5.03 & 0.50  & 0.36 & {\textbf{0.11}} & 0.30  &  {\textbf{0.11}}  & 0.47 & {\textbf{0.17}} & 0.37  &  0.18\\
4  & 20  & 04 & 0.22  & 0.10 & 9.99 & 0.50  & 0.68 & {\textbf{0.11}}  & 0.57 &  {\textbf{0.11}}  & 0.86 & {\textbf{0.17}} & 0.68  & {\textbf{0.17}}\\
5  & 05  & 04 & 0.05  & 0.10 & 0.29 & 0.50  & 0.14 & 0.11 & 0.13 & {\textbf{0.08}} &  0.16 & 0.18 & 0.15  & {\textbf{0.13}}\\	 
6  & 10  & 10 & 0.11  & 0.25 & 5.03 & 4.68  & 0.36 & 0.32 & 0.34  &   {\textbf{0.25}}  & 0.47 & 0.55 & 0.49  &{\textbf{0.43}}\\
7  & 20  & 20 & 0.22  & 0.50 & 9.99 & 13.12  & 0.68 & 0.54 & 0.64  &  {\textbf{0.46}}  & 0.85 & 0.85 & 0.78  &  {\textbf{0.73}}\\ \hline
\end{tabulary}
\end{table}

\begin{table}
\centering
\caption{\label{tab:results_largescales}
NRMSEs of large and small scales. Notations are explained in Table.~\ref{tab:results}.}
\vspace{.5cm}
\begin{tabulary}{0.6\textwidth}{|c|cccc|cccc|}
\hline
\multicolumn{1}{|c|}{}&\multicolumn{4}{c|}{$\overline{\epsilon}$}&\multicolumn{4}{c|}{$\epsilon_{max}$}\\ \hline
{Case} & {$\varmathbb{I}_s\boldsymbol{y}$} & {$\varmathbb{I}_t\boldsymbol{x}$} & {LSE} & {Fusion} & {$\varmathbb{I}_s\boldsymbol{y}$} & {$\varmathbb{I}_t\boldsymbol{x}$} & {LSE} & {Fusion}\\
\hline
\multicolumn{9}{|c|}{Large scales reconstruction} \\ \hline
5  & 0.08 & 0.09 & 0.10  & {\textbf{0.05}} &  {\textbf{0.06}} & 0.15 & 0.11  & 0.07\\
6  & 0.24 & 0.25 & 0.24  &   {\textbf{0.15}}  & 0.22 & 0.45 & 0.28  & {\textbf{0.21}}\\
7  & 0.56 & 0.36 & 0.51 &  {\textbf{0.30}}  & 0.60 & 0.66 & 0.60 &  {\textbf{0.46}}\\ \hline
\multicolumn{9}{|c|}{Small scales reconstruction} \\ \hline
5  & 0.98 & 0.56 & 0.73  & {\textbf{0.55}} &  0.98 & 0.86 & 0.89  & {\textbf{0.81}}\\
6  & 0.98 & 0.81 & 0.90  &  {\textbf{0.70}}  & 0.97 & 1.15 & 1.07  & {\textbf{0.92}}\\
7  & 0.99 & 0.92 & 0.95  &  {\textbf{0.78}}  & 0.93 & 1.08 & 0.90  &  {\textbf{0.86}}\\ \hline
\end{tabulary}
\end{table}
\end{landscape}
\section{\label{sec:numerical_experiments} Numerical experiments}
Section ~\ref{subsec:database} describes the DNS database used to test the model. Section ~\ref{subsec:other_reconstruction_methods} discusses other reconstruction methods for comparison. Section ~\ref{subsec:results} presents results of the fusion model in various cases. 

\subsection{\label{subsec:database} DNS database}
DNS database of a turbulent wall-bounded flow is used to test the model. This simulation uses the numerical procedure described in \cite{marquillie2008direct}.
The flow is at a Reynolds number $ Re_{\tau}=550 $ based on the friction velocity. Cartesian coordinates of the simulation in space are $ (x,y,z) $ for streamwise, vertical and spanwise directions respectively. The domain size $ L_x \times L_y \times L_z $ normalized by half the channel height $ H $ is $ 2\pi \times 2 \times \pi $. Fully resolved fluctuating streamwise velocities in a plane normal to the flow direction are considered as HTHS data. This data includes $ P=10000 $ snapshots at spatial resolution of $ N=288 \times 257 $ and at sampling frequency of 40 Hz. Sparse LTHS and HTLS measurements are subsampled from HTHS data to learn the fusion model. HTHS is used as the ground truth to estimate reconstruction errors. The extension to spanwise and vertical velocity components follows the same procedure.

Various cases are investigated. The subsampling ratios $ \sqrt{N/M} $ applied in each direction of space are 5, 10 and 20. These ratios correspond to a number $ M $ of HTLS sensors  of $ 51 \times 57 $, $ 26 \times 29 $ and $ 13 \times 15 $ respectively. Each ratio has a spacing between two successive HTLS points in spanwise and vertical directions of $ \Delta z $ and $ \Delta y $. Subsampling ratios $ P/Q $ in time are 4 ($ Q= 2500$), 10 ($ Q=1000 $) and 20 ($ Q=500 $). Each ratio, both in space and time, corresponds to a certain amount of energy loss. This is essentially the energy of small scales separated from large scales by a low pass filter $ \varmathbb{L} $. Here $ \varmathbb{L} $ is the $ 5^{th}-$order least square spline filter, either temporal 1D ($ \varmathbb{L}_t $) or spatial 2D ($ \varmathbb{L}_s $), using measurements as knots. This spline filter has the advantages of a sharp cutoff response and finite support. The energy loss is defined by comparing the filtered field $ \varmathbb{L}\boldsymbol{z} $ and the original field $ \boldsymbol{z} $: 
\begin{equation}
	\Delta\kappa=\frac{\sqrt{\sum\limits_{j\in \varmathbb{J}} \boldsymbol{z}_j^2}-\sqrt{\sum\limits_{j\in \varmathbb{J}}{[\varmathbb{L}\boldsymbol{z}]_j^2}}}{\sqrt{\sum\limits_{j\in \varmathbb{J}}\boldsymbol{z}_j^2}}
	\label{eq:RMS_losses}
\end{equation} 
where $ \varmathbb{J}$ is the considered set of points. Table~\ref{tab:results} gathers the energy loss in time ($ \Delta\kappa_t $) and in space ($ \Delta\kappa_s $) estimated with $ \varmathbb{L}_t $ and $ \varmathbb{L}_s $ respectively. The set $ \varmathbb{J}$ contains all points at $ y/H=1 $.

\subsection{\label{subsec:other_reconstruction_methods} Other methods for comparison}
Other reconstruction methods are used for comparison with the present model.

\textbf{Cubic spline interpolation}: Interpolation techniques reconstruct HTHS velocities from either LTHS or HTLS measurements independently, i.e. $ \boldsymbol{x} \longmapsto \hat{\boldsymbol{z}}=\varmathbb{I}_t \boldsymbol{x} $ or $ \boldsymbol{y} \longmapsto \hat{\boldsymbol{z}}=\varmathbb{I}_s \boldsymbol{y}$. The cubic spline interpolations \cite{unser1999splines}, either 1D in time or 2D in space, are used. These interpolations are by Matlab built-in functions, which follow the algorithm in Ref.~\cite{de1978practical}.

\textbf{Linear Stochastic Estimation}: 
LSE estimates $ \hat{\boldsymbol{z}} $ as a linear combination of measurements. Coefficients are estimated from the measurements by solving a system of linear equations to minimize the mean square errors of reconstructed fields. Refs.~\cite{adrian1979conditional,adrian1992stochastic} describe the physical interpretations of this procedure. This section derives the model differently \cite{hastie2009elements,bishop2006pattern} but in accordance with turbulence literature.

Matrix forms $ X $, $ Y $ and $ Z $ described in Section ~\ref{subsec:statistical_parameters_estimation} are used to build the LSE model. Let $Y_s = \varmathbb{S}_t Y$ of size $ Q \times M $ denote a part of $ Y $ subsampled at the same instants as $ X $. LSE model finds the optimal matrix $B$ of size $ N \times M $ that minimizes the residual sum of squared errors:
\begin{equation}
B=\argmin_B \: \Vert Y_sB-X\Vert^2_2
\label{eq:LSE1}
\end{equation}
Let set the gradient of this residual sum to zero:
\begin{equation}
\frac{\partial \Vert Y_sB-X\Vert^2_2}{\partial B}= Y_s^T(Y_sB-X)=0
\label{eq:LSE2}
\end{equation}
the optimal $ B $ is obtained as:
\begin{equation}
B=\left( Y_s^TY_s \right)^{-1} Y_s^TX
\label{eq:LSE3}
\end{equation}
Eq.~(\ref{eq:LSE3}) requires the inversion of $ (Y_s^TY_s) $ that can be singular, leading to a high variance model with large coefficients. A small change of predictors $Y$ then can lead to a very different reconstruction of $ Z $, causing model's instability. Tikhonov regularization \cite{cordier2010calibration}, well-known in machine learning problems as L2 penalty or Ridge Regression \cite{hastie2009elements,bishop2006pattern}, can be used as a remedy. It aims to solve this ill-posed problem by imposing a L2 penalty term on the residual sum of errors. The optimization problem (\ref{eq:LSE1}) becomes:
\begin{equation}
B=\argmin_B \: \Vert Y_sB-X\Vert^2_2 +\lambda\Vert B \Vert^2_2
\label{eq:RR1}
\end{equation}
Setting the gradient of the cost function (for $ \lambda>0 $) to zero:
\begin{equation}
\frac{\partial \left( \Vert Y_sB-X\Vert^2_2 +\lambda\Vert B \Vert^2_2 \right)}{\partial B}= Y_s^T(Y_sB-X) + \lambda B=0
\label{eq:RR2}
\end{equation}
the closed form of $ B $ is:
\begin{equation}
B=\left( Y_s^TY_s+\lambda I \right)^{-1}  Y_s^TX 
\label{eq:RR3}
\end{equation}
The regularization parameter $ \lambda $ can be optimized by ten-fold cross-validation \cite{geisser1993predictive}. The fully resolved field of $ Z $ is then estimated using these coefficients:
\begin{equation}
Z=YB
\label{eq:RR4}
\end{equation}
Matrix $B$ encodes the predictor of $Z$ knowing $Y$ learnt from the joint observation of $X$ and $Y$. A completely analogous procedure can be used switching the roles of $ X $ and $ Y $.

\subsection{\label{subsec:results} Results}
\subsubsection{\label{subsubsec:impacts_of_subsampling_ratios} Impact of subsampling ratios}
%\begin{figure}
%\includegraphics[width=0.8\columnwidth]{./figures/experimentalsetup/element_block.png}
%\caption{\label{fig:element_block} Sketch of an element block with local coordinates $ (\alpha,\beta,\tau) $. LTHS time steps are at $ \tau/\delta t=0 $ and $ \tau/\delta t=P/Q $. HTLS measurements are represented by red dots and LTHS measurements by black ones. }
%\end{figure}
The fusion model uses Eq.~(\ref{eq:MAP_simplified}) to reconstruct fully resolved velocities $ \hat{\boldsymbol{z}} $ in various cases. Reconstructed fields are compared with the original DNS via the Normalized Root Mean Square Error (NRMSE):
\begin{equation}
NRMSE = \left(\frac{ \sum\limits_{j\in \varmathbb{J}}(\hat{\boldsymbol{z}}_j-\boldsymbol{z}_j)^2}{\sum\limits_{j\in \varmathbb{J}}\boldsymbol{z}_j^2}\right)^{1/2}
\label{eq:NRMSE}
\end{equation}
where $ \varmathbb{J} $ is the considered set of points used to estimate the error. The field $\boldsymbol{z}$ is more or less difficult to estimate depending on the considered instant and position with respect to available measurements. To qualify, two types of NRMSE, the mean NRMSE $ \overline{\epsilon} $ and the maximum NRMSE $ \epsilon_{max} $, are estimated. $ \overline{\epsilon} $ is estimated over $ \varmathbb{J}$ including all space-time positions in the outer region of $ y/H \in [0.25,1.75] $, where the flow is approximately homogeneous. It represents how far the reconstructed field departs from ground truth in order to evaluate reconstruction accuracy. 
%To estimate $ \epsilon_{max} $, let define element block with local coordinates as in Fig.~(\ref{fig:element_block}). 
$ \epsilon_{max} $ is estimated using all blocks (in time and in spanwise directions) bounded by HTLS sensors at $ y/H= 0.94 $ and $ y/H=1.06 $, see Fig.~\ref{fig:element_block}. The set $ \varmathbb{J} $ includes centers at local coordinates  $ (\Delta y/2, \Delta z/2, P \delta t/2Q) $ of all blocks. $ \overline{\epsilon} $ and $ \epsilon_{max} $ of $ \varmathbb{I}_s\boldsymbol{y} $, $ \varmathbb{I}_t\boldsymbol{x} $ and LSE reconstruction are also estimated for comparison. 

Table~\ref{tab:results} describes 7 cases with their settings and reconstruction errors. In cases 1 and 2, the energy losses due to subsampling in time are much higher than in space, and vice-versa in cases 3 and 4. The model gives similar errors compared to the best interpolation, with smaller $ \overline{\epsilon} $ and comparable $ \epsilon_{max} $. In cases 5 to 7, the losses are due to both the subsamplings in space and time in a balanced manner. The proposed model reduces $ \overline{\epsilon} $ by 15$ \% $ to 30$ \% $ and $ \epsilon_{max} $ by 10$ \% $ to 20$ \% $ compared to the best of other methods. 
 
Improvements are expected from the weighted average in Eq.~(\ref{eq:MAP_simplified}). The present model uses variances $ \boldsymbol{\sigma}^2_s(i) $ and $\boldsymbol{\sigma}^2_t(i) $ as parameters of the flow's physics, and $ \varmathbb{I}_t\boldsymbol{x} $ and $ \varmathbb{I}_s\boldsymbol{y} $ as the specific flow information. It imposes the reconstruction to be consistent with measurements at nearby positions and proposes compromise estimates elsewhere. Simple interpolations use either HTLS or LTHS measurements only, losing information from the other source. LSE learns its coefficients from both measurements but inherits the limitations of the conditional averaging. 

\subsubsection{\label{subsubsec:large_and_small_scales_reconstruction} Large and small scales reconstruction}

In cases 1 to 4, the fusion model performs as the best interpolation with small improvements. This is expected since one measurement of HTLS or LTHS is much better resolved than the other. Cases 5 to 7  are the most interesting since energy losses due to subsampling in space and time are comparable. The model brings complementary information from both measurements and improves the reconstruction.

We study reconstructions of large and small scales in details for these three cases. Spatial 2D filters $ \varmathbb{L}_s $ (see Section \ref{subsec:database}) are used to separate large scales from small scales. These filters take HTLS points as knots to have a cutoff close to the Nyquist frequency. The reconstructed large scales by all methods are compared to the reference $ \varmathbb{L}_s\boldsymbol{z}$. Small scales are estimated using $\mathbf{I}-\varmathbb{L}_s $ where $ \mathbf{I} $ is the identity matrix. Table~\ref{tab:results_largescales} shows NRMSEs estimated using Eq.~(\ref{eq:NRMSE}) but normalized by the RMS of either $ \varmathbb{L}_s\boldsymbol{z}$ or $(\mathbf{I}-\varmathbb{L}_s)\boldsymbol{z} $. 

The fusion model recovers part of small scales from complementary measurements. It gives the lowest $ \overline{\epsilon} $ and $ \epsilon_{max} $ of small scales reconstruction in all cases. It also better reconstructs large scales than other methods. For large scales, $ \epsilon_{max} $ remains the same in case 5 of small subsampling ratios and improves significantly in cases 6 and 7 of high ratios, with $ \epsilon_{max} $ reduced by 5 $ \% $ and 25 $ \% $ respectively, and $ \overline{\epsilon} $ by 20$ \% $ to 40$ \% $ compared to the best of other methods.

\subsubsection{\label{model_performance_analysis} Model performance analysis}
\begin{figure}
\centering
\subfigure[]{\includegraphics[height = 5.5cm]{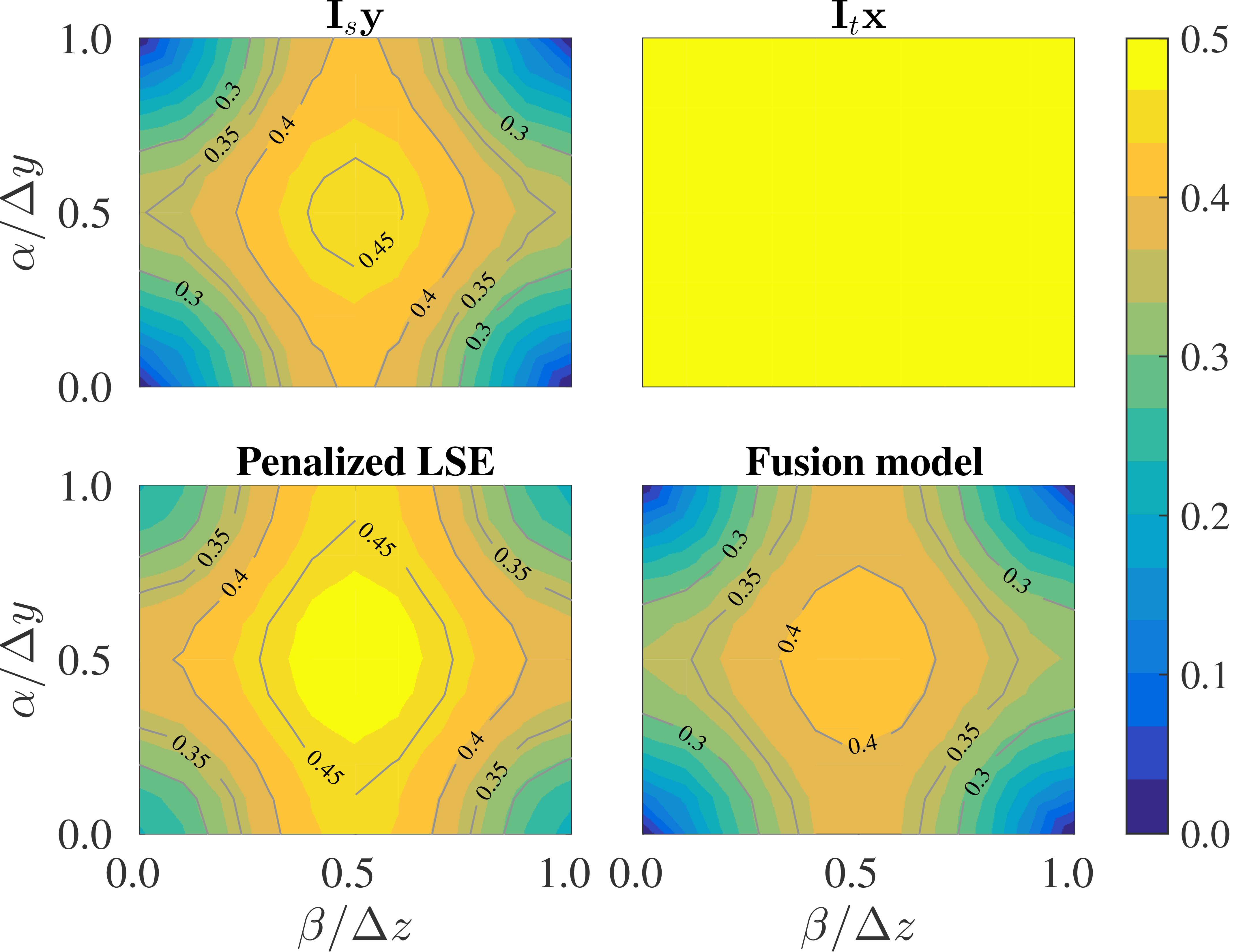}
\label{fig:error_MAP_newOM1_outer_a}}
\quad
\subfigure[]{\includegraphics[height = 5.5cm]{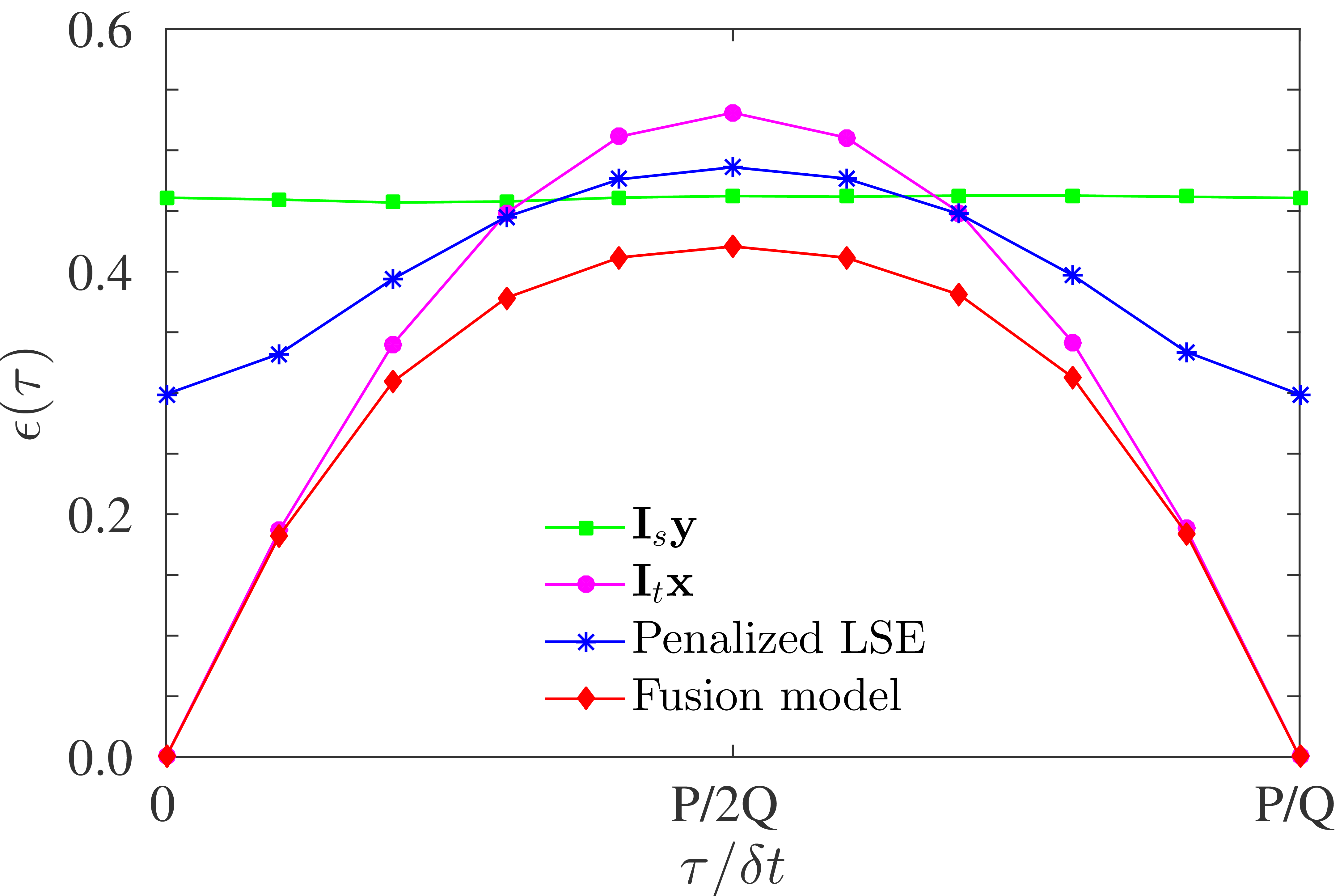}
\label{fig:error_MAP_newOM1_outer_b}}

\caption{\label{fig:error_MAP_newOM1_outer} NRMSEs between reference and reconstructed streamwise velocities by all methods as: (a) functions of spatial coordinates in an element block at the most difficult instant, i.e. at $ (\alpha,\beta,P \delta t/2Q) $; (b) functions of time distances from the previous LTHS instant at the most difficult spatial location, i.e. at $ (\Delta y/2,\Delta z/2,\tau) $.}
\end{figure}

\begin{figure}
\includegraphics[width=\textwidth]{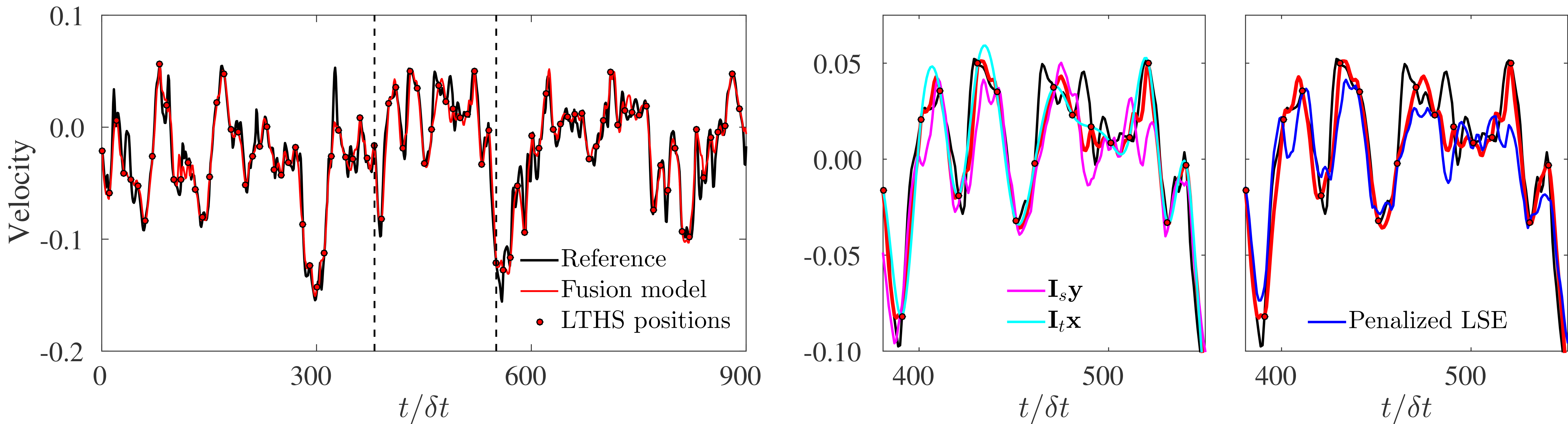}
\caption{\label{fig:improper_point_spacespacing_10_timespacing_10_yid129_zid149} A time evolution of fluctuating streamwise velocity at $ y/H=1 $ and $ z/H=0 $, the centers of all such $ (\alpha,\beta) $ planes in Fig.~(\ref{fig:element_block}).}
\end{figure}

\begin{figure}
\centering
\includegraphics[width=0.7\textwidth]{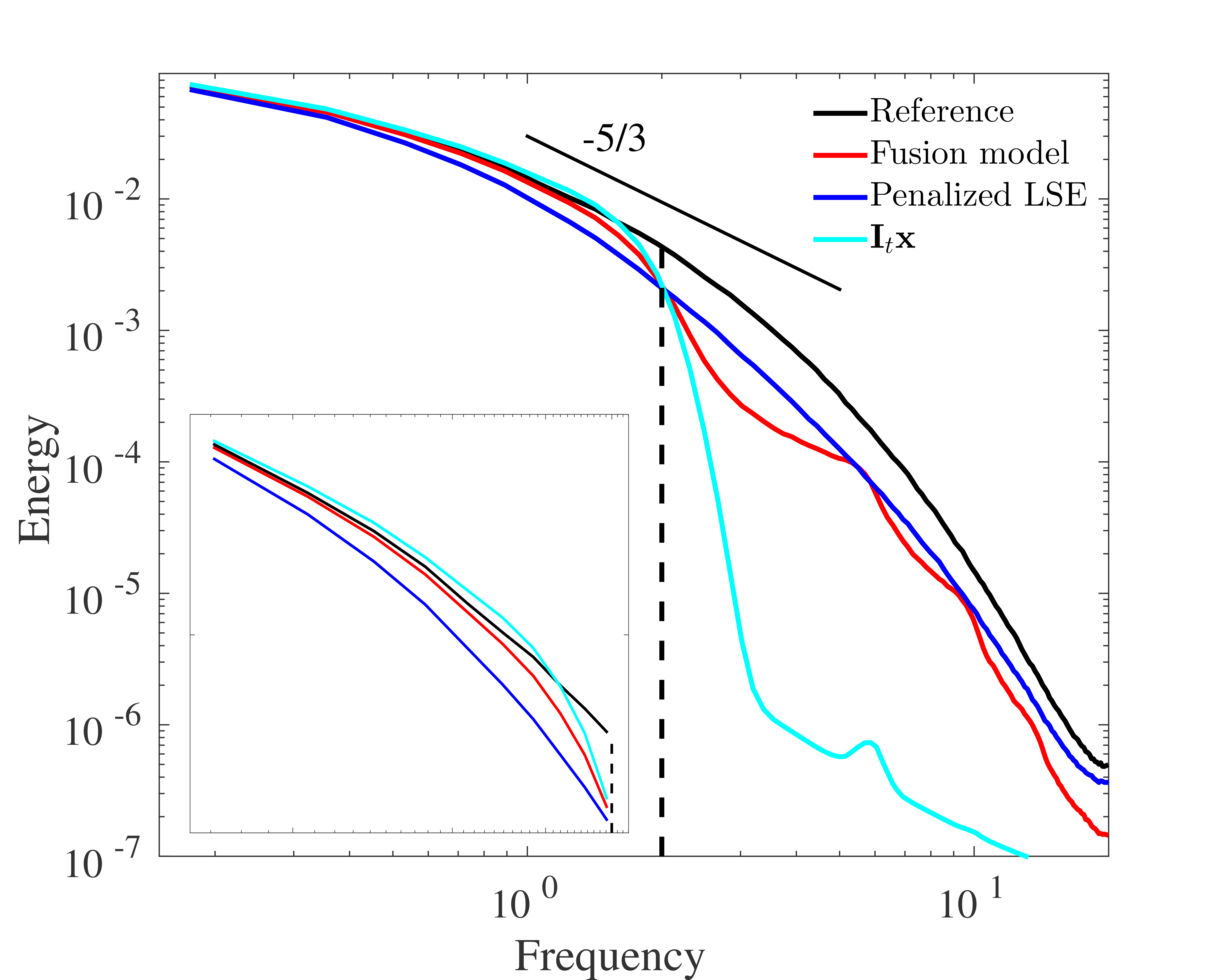}
\caption{\label{fig:improper_sspacing_10_tspacing_10_spectrum_time} Spectra of the fluctuating velocity in Fig.~\ref{fig:improper_point_spacespacing_10_timespacing_10_yid129_zid149} .}
\end{figure}

\begin{figure}
\centering
\includegraphics[width=\textwidth]{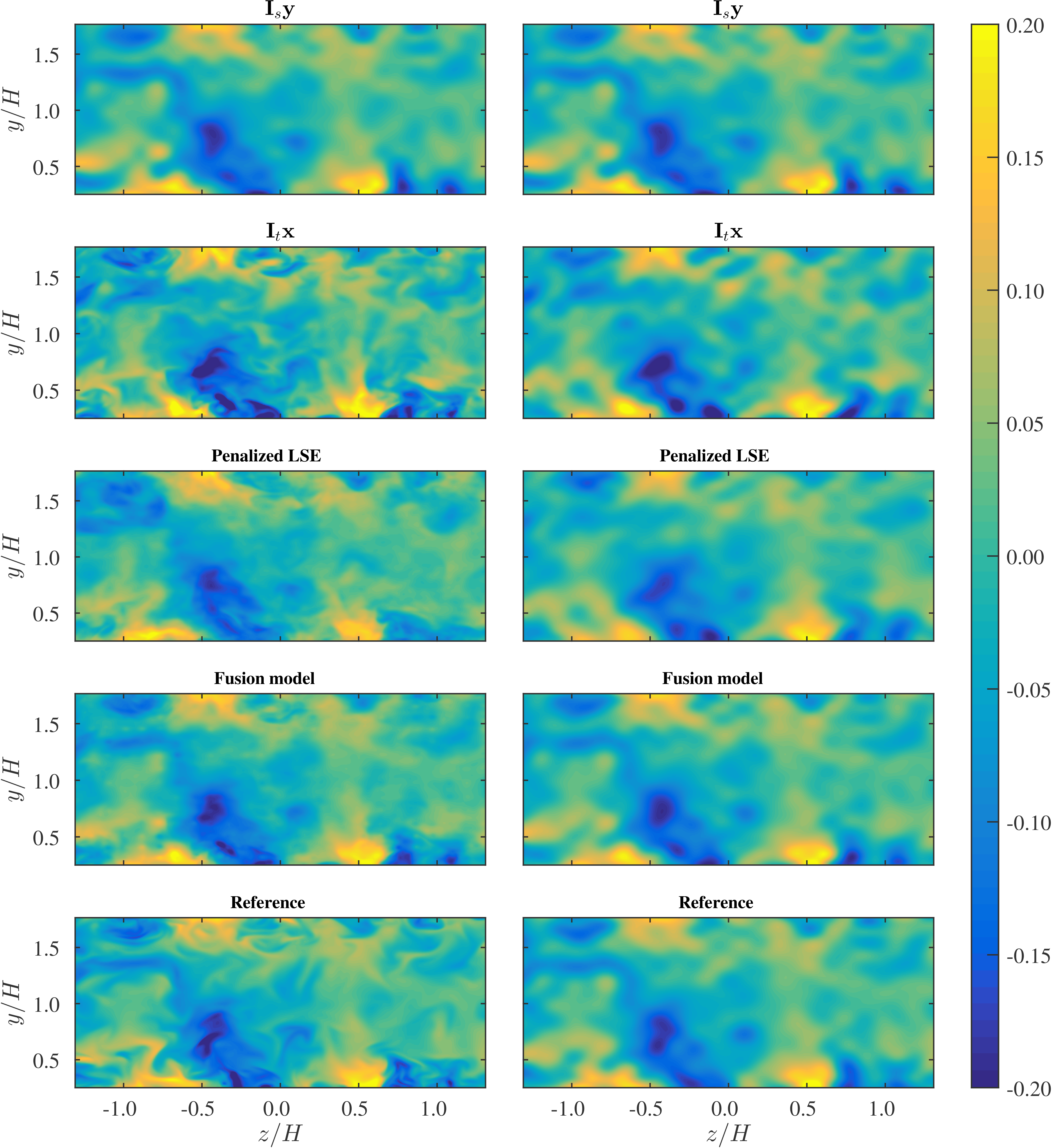}
\caption{\label{fig:improper_outer_spacespacing_10_timespacing_10_subplots_all_t006} A sample snapshot of fluctuating streamwise velocity at one of the most difficult instant to estimate (in the middle of two LTHS time steps): Reconstruction of all scales (left) and large scales only (right). The figure is better viewed on screen.}
\end{figure}

We focus on case 6 for a model performance analysis. This case has about 5 $ \% $ energy losses due to both time and space subsamplings, which are critical to highlight interests of the present approach. The model reduces  $ \overline{\epsilon} $  and $ \epsilon_{max} $ by 25 $ \% $ and 35 $ \% $ respectively for all scales reconstruction, 10 $ \% $ and 5 $ \% $ for large scales reconstruction. 

To analyze reconstructions in space, Fig.~\ref{fig:error_MAP_newOM1_outer}(a) shows spatial NRMSE maps by all methods as functions of local coordinates $ (\alpha,\beta) $. For each $ (\alpha,\beta) $, NRMSE is estimated using Eq.~(\ref{eq:NRMSE}), where $ \varmathbb{J} $ includes points at $ (\alpha,\beta,P \delta t/2Q) $ of all blocks used to estimate $ \epsilon_{max} $ (see Section~\ref{subsubsec:impacts_of_subsampling_ratios}). For all methods, NRMSEs are small close to the four HTLS positions in the corners and increase when approaching the center. Time interpolation behaves differently since its errors are independent of spatial coordinates. The fusion model yields the smallest errors at all positions. It improves significantly near the center compared to spatial interpolation, the best of other methods. 

To analyze reconstructions in time, Fig.~\ref{fig:error_MAP_newOM1_outer}(b) shows the NRMSE curves by all methods as functions of distances $ \tau $ from the previous LTHS time step. For each $ \tau $, NRMSE is estimated using $ \varmathbb{J}$ including points at local coordinates $ (\Delta y/2,\Delta z/2, \tau) $ of all blocks used to estimate $ \epsilon_{max} $. NRMSEs are small close to the LTHS measurements ($ \tau/\delta t=0 $ and $ \tau/\delta t=P/Q $) and increase when moving toward the middle ($ \tau/\delta t=P/2Q $). Spatial interpolation are different with NRMSEs independent of time. The fusion model yields the minimum errors at all time steps. Even in the middle of two LTHS instants, the maximum fusion error remains significantly lower than that of all other methods.   

Fig.~\ref{fig:improper_point_spacespacing_10_timespacing_10_yid129_zid149} shows a time evolution of the point at $ y/H=1 $ and $ z/H=0 $ ($ \alpha=\Delta y /2 $ and $ \beta = \Delta z/2 $ in local coordinates), the most remote from its neighboring HTLS sensors. A good agreement between fused and reference velocity is still obtained. A zoom-in period is shown also for detailed comparisons with other methods. While time interpolation captures only low frequencies, spatial interpolation generates high frequencies but weakly correlated with the truth. The fusion model proposes a good compromise to improve both large and small scales reconstruction. It also captures detailed peaks much better than LSE, since LSE smooths these small scales out by minimizing the mean square errors. 

Fig.~\ref{fig:improper_sspacing_10_tspacing_10_spectrum_time} compares temporal spectra of above evolutions. Time interpolation fails to estimate the signal at higher frequencies than a certain cutoff. LSE keeps both large and small scales, but the loss of large scale energy is critical. This loss is highlighted in the zoom-in picture of low frequencies spectral. The present model improves the estimation at both low and high frequencies.   

Fig.~\ref{fig:improper_outer_spacespacing_10_timespacing_10_subplots_all_t006} compares reconstructed snapshots by different methods. This snapshot is at the most remote instant from its two neighboring LTHS time steps. The model reconstructs correctly the velocity field with more flow details than spatial interpolation. It also recovers better large scales than LSE and time interpolation methods.

%%%%%% CONCLUSION
\section{\label{sec:conclusions} Conclusions}
This work proposes a Bayesian fusion model using a MAP estimate to reconstruct high resolution velocities of a turbulent channel flow from low resolution measurements in space and time. It searches for the most probable field given available measurements. This approach yields a simple but efficient weighted average formula in Eq.~(\ref{eq:MAP_simplified}). Weighting coefficients are learnt from measurements and encode the pysics of the flow. The informed fusion of information from available measurements improves the interpolation of large scales and recovers details at small scales.

Numerical experiments using a DNS database of a turbulent wall-bounded flow at a moderate Reynolds number illustrate the efficiency and robustness of the proposed method. Low resolution measurements are extracted to learn model parameters, while original data are used as the ground truth to estimate reconstruction errors. The model is tested in various cases with different subsampling ratios. Results are compared to more standard methods such as cubic spline interpolation and penalized LSE. Bayesian fusion always produces the most accurate reconstruction. The best results are obtained when missing spatial and temporal information are of the same order of magnitude. In these cases, it provides a better large scale reconstruction while a certain amount of small scale details are also recovered. The search for an even more accurate fusion and super-resolution method is the subject of ongoing work.

%\bibliographystyle{unsrt}
%\bibliography{nguyen}

\end{document}